# Custom RFID Location Simulator

Hakan Yılmaz, Osman Nacar, Özgür Sezgin, Erkan Bostancı, Mehmet Serdar Güzel
SAAT Laboratory
Computer Engineering Department, Ankara University
Ankara, Turkey
ylmzhakann@gmail.com, oncr@outlook.com, ozgurr05@gmail.com, ebostanci@ankara.edu.tr, mguzel@ankara.edu.tr

*Abstract* — **Radio frequency identification (RFID), The real-time location of objects and ability to track motion provide a wide range of useful applications in areas such as safety, security and supply chain. In recent years, radio frequency identification technology has moved from obscurity into mainstream applications that help speed the handling of manufactured goods and materials. RFID enables identification from a distance, and unlike earlier bar-code technology, it does so without requiring a line of sight. In this paper, the author introduces the principles of RFID, discusses its primary technologies and applications.**

*Keywords—RFID technologies, Trilateration Formula, Signal Permeability Coefficient, RFID tags ,Tracking, Passive RFID tags, Active RFID tags,*

## I. INTRODUCTION

Wireless technologies have great proposition in our daily lives. Wi-Fi, Bluetooth, NFC, IrDA and ZigBee technologies are used in many areas of wireless communication. File sending / receiving, display speech, device sharing and many other applications can be used with wireless systems. RFID technology, which was developed for use in war technology in the mid-20th century, also took place in wireless systems. The rapid development of this technology, which was developed to describe physical objects, also made significant contributions to production activities. Considering the development, it is seen as a promising technology for the future. RFID is a technology that can be applied in many areas such as supply chains, health, animal husbandry, education, library, security.

This paper provides information on what RFID technology is. The basic components used in RFID systems and how they are used, their properties, their purpose of use are mentioned. After informing about RFID technology, information about RFID simulation study has been given. In this information, how the RFID simulation study works, the usage of the interface is given. The formulas used in the study are mentioned and the working logic of the formulas is explained. For the remainder of the article, the RFID components will be presented in Part II, and the development environment in Part III. The experimental results will be presented in Part IV and the results in Part V.

## II. RFID TECHNOLOGIES BACKGROUND

*1) Basic Components Of An Rfid System*

Basic RFID systems consist of three hardware components and three types software components. These hardware components are tag, antenna and reader and software components are firmware, middleware and application software.

*1.1) Hardware Components*

First hardware component is tag. The tag is a chip set in which information about the object is stored and components that contain an antenna for communicating with the reader. They use radio frequency signals to communicate with the reader. Each label has a unique identifier number. [3]

Second hardware component is antenna. RFID antennas provides communication between reader-reader or reader-tag. The main purpose of an antenna is to radiate the electromagnetic waves produced by the reader, and in the same manner, receives radio frequency signals from the tags.

The last hardware component is the reader. RFID readers communicate with tags. The RFID reader is a hardware that can read the tag information by receiving the signal from the back of the RFID tag and can write new information of the tag by spreading the signal through the radio frequency.[4][5][6]

*1.2) Software Components*

A few different types of software are common components of most RFID systems firmware, middleware, and application software. Though all of these components are technically software, their individual functions differentiate them into one of the aforementioned three categories.

First software component is application software. Application software is any set of machine-readable instructions that directs a computer's processor to perform specific operations. Thousands of software applications are accessed daily by end-users, ranging from apps on our phones, to some more specialized applications such as software built for accessing and analyzing data collected by RFID systems.

Second software component is firmware. Firmware controls the operation of the device on which it is hosted and does not typically initiate communication with external devices, such as PCs. Device firmware may be upgraded periodically to fix bugs and to add new functionality to the hardware component.

Third software component is middleware. Middleware is a piece of software that usually runs in the background. A common use of RFID middleware is a service that communicates with and controls RFID readers in order to gather data, which then may be analyzed and stored in a database for consumption by a different user-facing application.

*2) Rfid Systems*

RFID systems comprise data-collecting readers and data-providing transponders, or tags, which are affixed to the physical objects to be tracked. In operation, readers query tags by transmitting an interrogatory signal on the system's operating frequency, and tags respond by transmitting data stored on the tag. To accomplish this exchange, readers include an RF transceiver, antenna, and controller capable of managing the communications protocol and data. Tags at a minimum consist of an antenna and tag chip, which contains an RF analog-front end with modulation circuitry, control logic, and memory. Tag capabilities and configurations vary widely, ranging from simple factory-programmed, read-only tags to field-rewritable, secure-communications tags.

RFID tags are available as strips, chips, swallowable capsules, and even embedded in hardware like screws for attachment to nearly any object. Conventional passive RFID tags contain no internal power source, but instead harvest power from the RF signal transmitted by the RFID reader system. For these tags, the reader and tag antennas are inductively coupled, forming a transformer with the reader as primary and tag as secondary, separated by a sizeable air gap. RFID readers provide sufficient power in the transmitted RF carrier signal to energize the tag circuitry through the coupled tag antenna. In addition, some RFID systems use the carrier to provide a clock source to tags, which divide the carrier frequency down to the clock frequency required for tag circuitry.

RFID systems can be either active or passive. Passive tags are powered by electromagnetic induction via the RFID reader but Active RFID tags contain a battery and periodically transmit information with much greater range than passive tags.

The reader communicates with the label via electromagnetic waves. Electromagnetic waves transmitted by the reader meet with the chip as an energy, move it into action, and transfer data from the tag to the reader. All this happens at a certain distance, without wirelessly.

## III. DEVELOPED SYSTEM

The aim of this study is real time object tracking. RFID Simulation works with four reader and tags. Tags are affixed to objects and readers follow objects with the help of these tags. The distance between the tags and the reader is determined by the radio signals sent by the readers and the position of the object is determined using the Trilateration formulas.

Simulation study, it is aimed to find the position of instant moving objects by using 4 different readers. The intersection points of the signals sent by the readers to the positions of the tags in the study are associated and the positions of the tags are determined using the Trilateration formula. When the position of the tags in the study is detected, The actual position is compared with the position found by the players. The actual position is shown in the graph with the position found depending on the instantaneous movement of the object.

The Trilateration Formula allows us to find the intersection point when it is known that 3 circles with known radii intersect at a single point. Today, this formula is behind the finding of the instant position of the technologies like GPS. 3 It is possible to determine the positions of the signals coming from different satellites.

It is known that the distance from Satellite 1 to the reception signal is not enough to determine the location of this information. Then, an object signal is sent from another satellite in a different position and the distance is found. As shown in Fig. 1.

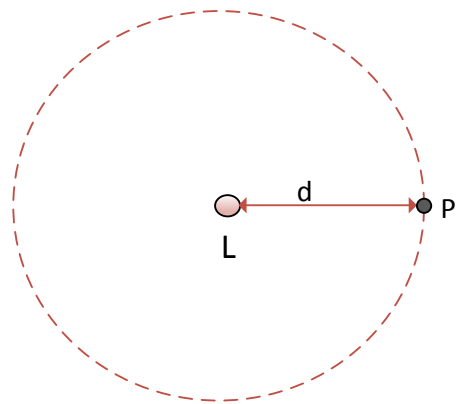

*Fig. 1 Communication with Satellite 1*

Satisfied with the information from Satellite 2's conformity, there is a distance from that satellite, but only this information is not sufficient and when we combine it with the information from Satellite 1's conformity, it may come to us that these two suits may be at 2 points when the intersection point of the circles drawn to the object is taken away and information from another satellite is requested.

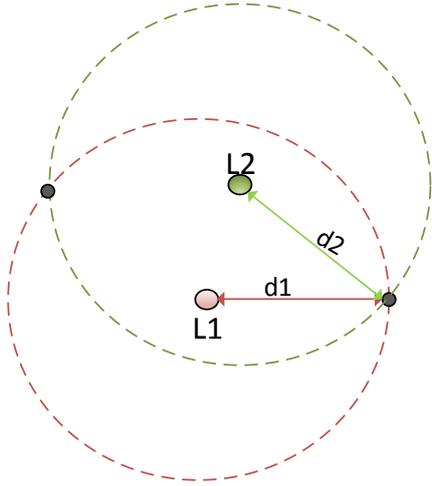

*Fig. 2 Communication with Satellite 1,2*

With the information and distance from satellite 3, it is now clear that there is only one point of intersection and that the position of the object is determined in this way. In the RFID Simulation study, the coordinates of the 4 different readers are known and the distances to the object are determined according to these coordinates. The 3 readers with the closest object are located at the intersection point with the Trilateration formula and the location is determined.

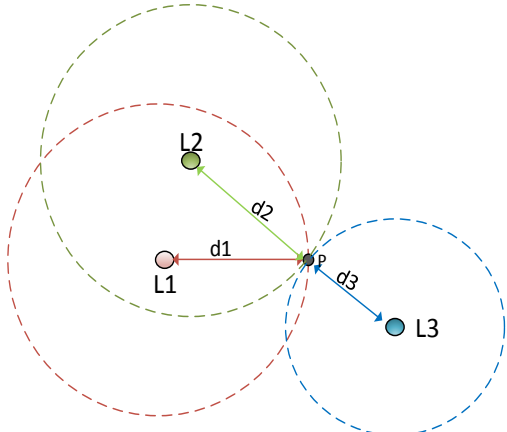

*Fig. 3 Communication with Satellite 1,2,3*

## IV. EXPERIMENTAL RESULTS

Trilateration formula uses the distance equation. If your unknown point is $(x,y)$, your known points are $(x_i,y_i)$ which are distances $r_i$ from your unknown point, then you get three equations:

$$(x - x_1)^2 + (y - y_1)^2 = r_1^2$$
$$(x - x_2)^2 + (y - y_2)^2 = r_2^2 \quad (1)$$
$$(x - x_3)^2 + (y - y_3)^2 = r_3^2$$

We can expand out the squares in each one:

$$x^2 - 2x_1 x + x_1^2 + y^2 - 2y_1 y + y_1^2 = r_1^2$$
$$x^2 - 2x_2 x + x_2^2 + y^2 - 2y_2 y + y_2^2 = r_2^2 \quad (2)$$
$$x^2 - 2x_3 x + x_3^2 + y^2 - 2y_3 y + y_3^2 = r_3^2$$

If we subtract the second equation from the first, we get

$$(-2x_1 + 2x_2)x + (-2y_1 + 2y_2)y = r_1^2 - r_2^2 - x_1^2 + x_2^2 - y_1^2 + y_2^2 \quad (3)$$

Likewise, subtracting the third equation from the second,

$$(-2x_2 + 2x_3)x + (-2y_2 + 2y_3)y = r_2^2 - r_3^2 - x_2^2 + x_3^2 - y_2^2 + y_3^2 \quad (4)$$

This is a system of two equations in two unknowns:

$$Ax + By = C$$
$$Dx + Ey = F$$

which has the solution:

$$x = \frac{CE - FB}{EA - BD} \qquad y = \frac{CD - AF}{BD - AE}$$

The permeability coefficients of each object are different from each other, and these permeability coefficients may vary depending on the thickness, brightness and type of objects.

In the study, the errors of position detection are shown by the signal losses which may be according to the permeability coefficients of the objects. The electrical permeability coefficients of the objects that can be determined are shown. Limestone with the most coefficient of limestone and styrofoam with the lowest coefficient of permeability are shown.

| Signal Permeability Coefficients of Objects [7] | | |
|---|---|---|
| **Material** | **GHZ** | **Thickness(cm)** |
| Tree structure | 2,19 | 2,07 |
| Thin Wall | 1,29 | 5,93 |
| Glass | 4,00 | 0,24 |
| Iron door | 2,07 | 4,45 |
| Ready-made panel wall | 2,44 | 4,17 |
| Plywood | 2,55 | 1,12 |
| Styrofoam | 1,12 | 9,91 |
| Brick | 4,20 | 15 |
| Concrete block | 2,30 | 30 |
| Limestone | 7,51 | 10 |

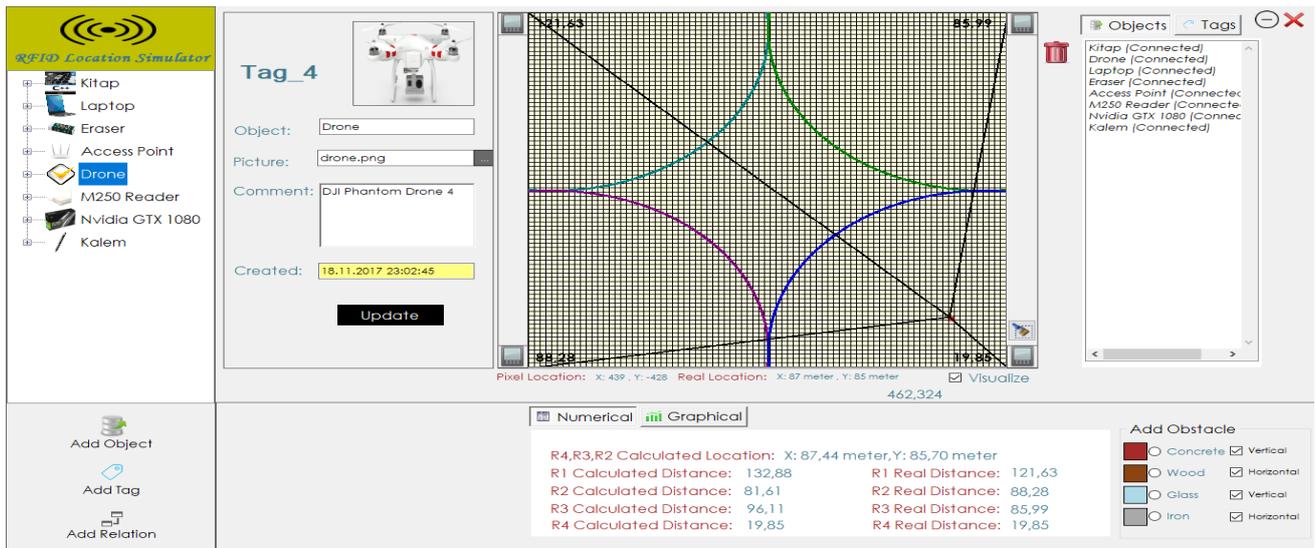

*Fig. 4 RFID Simulation Study*

After seeing the information of the selected object, we can update the information by clicking the update button and the distances from the drone are shown. When the position of the object is found, 3 of the closest to the object are used. From the distances written in the years, the ones to be used are selected and the places where the readers are located are considered as the centers and the selected 3 readers are considered as circles and the trilateration formula is used and the position of the object is determined.

Real Location and the coordinates of the actual location of the object are written in meters, the place where the object is located in Pixel Location and the 500x500 pixel screen. With Real Location, we are converting a 500x500 pixel area to 100x100 meters. In the lower part, we show the position calculated by the reader with the Numerical tab, which readers calculate and the distances of all readers to the object.

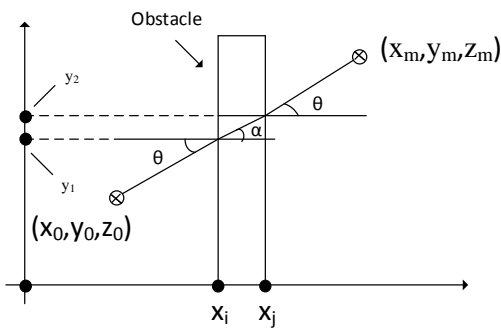

*Fig. 5 Calculation Technique of Signal Permeability of Objects*

It is explained how the signal changes when the obstacle between the objects and the readers is placed and how the calculations of this change are done. When the losses are calculated, the calculations are made taking into account the obstacle's gender and thickness.

To talk about the obstacles we use in our study, we used 4 different objects in the study: wall, glass, wood and iron.

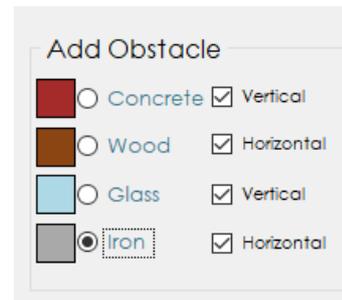

*Fig. 6 Add Obstacles Section*

We aimed to obtain object close to real data by calculating the signal losses according to the coefficients of permeability of objects by putting our object in two different ways, horizontally and vertically, between objects and 4 different coordinate readers.

Using this data, which is obtained from incorrectly calculated coordinates of the actual coordinates of the objects, the error graphs are plotted to provide the user with more meaningful data. As shown in Fig. 7.

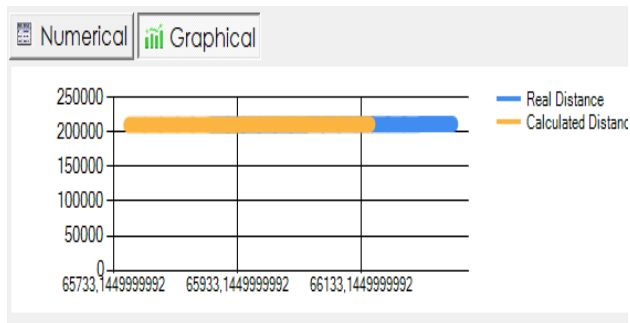

*Fig. 7 Graphical statistics section*

It shows how far the simulation works efficiently by presenting the actual distance and calculated distances of the objects to the readers instantly to the user.

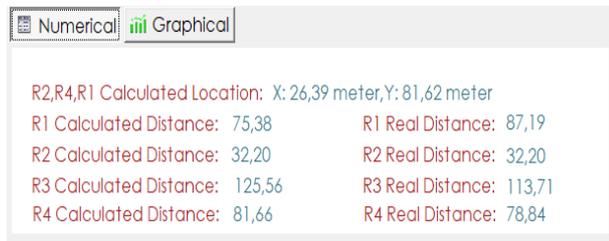

*Fig. 8 Calculated and Real Distance Area*

When the Visualize control button is pressed, the data of the backgrounds of the readers in the background of the study, such as the distance to the object, are displayed. The radius of the circles is the range of the readers. The ranges are assigned as 50 meters from the beginning of the program, but the distances can be changed when the readers are specifically clicked. The distance between the object and the reader is shown by the lines drawn from the object to the readers. It is the place where the end points of the correct parts drawn from each reader are objects.

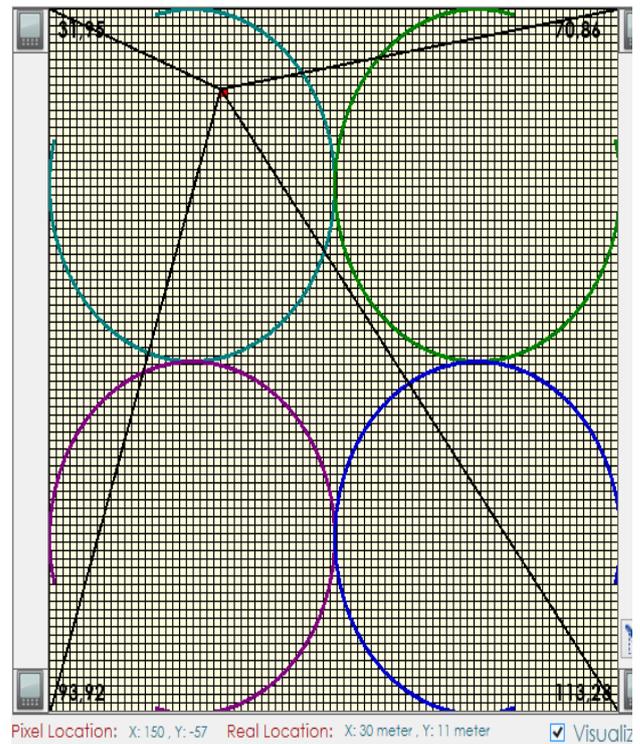

*Figure 9. Visualize Simulation Display*

The range defined by default for our readers is set at 50 meters, but these range distances can be changed by clicking on icons at the corner of the reader at the request of the user.

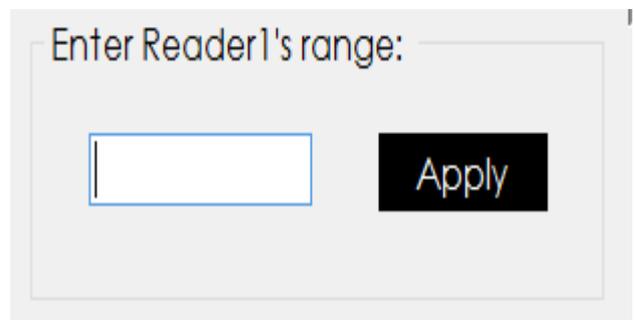

*Fig. 10 Enter Readers Range Section*

## V. CONCLUSION

In an object tracking system prepared with real hardware by RFID Simulation study, it is desired to show what hardware is needed for finding the position of the object, how these devices communicate and how accurate the system can work.

It has been learned what kind of equipment is needed in the systems requiring object tracking, how to communicate in order for the hardware to work correctly, and what formulas should be used for position detection.

It is shown in the graph that the results obtained in the study are how wrong the output of the results is in a real RFID system and the difference between these errors and the actual values is shown on the graph. It has been learned how the errors can arise and in what conditions the system can be used better.

This study can be used in systems such as inventory tracking systems, animal tracking systems, patient tracking systems, valuable goods tracking systems in jewellery, logistics tracking systems, etc. which require object tracking with real hardware (reader, tag, antenna) these errors can be eliminated by simulations and more accurate results can be obtained in real systems.